\documentclass[11pt]{article}
\pdfoutput=1
\usepackage[centertags]{amsmath}
\usepackage[square, comma, sort&compress,numbers]{natbib}
\usepackage{array,multirow}
\numberwithin{equation}{section}
\usepackage{amssymb,amsfonts}
\usepackage{graphicx}
\usepackage{color}
\usepackage{mathtools}

\usepackage{epsfig}
\usepackage{bbold}

\usepackage{wrapfig}
\usepackage{float}
\usepackage{soul}

\usepackage{tkz-euclide}

\usepackage{tikz,pgf}
\usetikzlibrary{shapes}
\usetikzlibrary{calc}
\usetikzlibrary{decorations.pathmorphing}
\usetikzlibrary{decorations.pathreplacing,shapes.misc}
\usetikzlibrary{positioning}
\usetikzlibrary{arrows}
\usetikzlibrary{decorations.markings}
\usetikzlibrary{shadings}
\usetkzobj{all}

\usetikzlibrary{intersections}

\def\be{\begin{equation}}
\def\ee{\end{equation}}
\def\ba{\begin{array}}
\def\ea{\end{array}}

\def\dps{\displaystyle}
\newcommand{\half}{\frac{1}{2}}

\def\1{\tilde{1}}
\def\2{\tilde{2}}
\def\3{\tilde{3}}

\newdimen\tableauside\tableauside=1.0ex
\newdimen\tableaurule\tableaurule=0.4pt
\newdimen\tableaustep
\def\phantomhrule#1{\hbox{\vbox to0pt{\hrule height\tableaurule
width#1\vss}}}
\def\phantomvrule#1{\vbox{\hbox to0pt{\vrule width\tableaurule
height#1\hss}}}
\def\sqr{\vbox{%
  \phantomhrule\tableaustep

\hbox{\phantomvrule\tableaustep\kern\tableaustep\phantomvrule\tableaustep}%
  \hbox{\vbox{\phantomhrule\tableauside}\kern-\tableaurule}}}
\def\squares#1{\hbox{\count0=#1\noindent\loop\sqr
  \advance\count0 by-1 \ifnum\count0>0\repeat}}
\def\tableau#1{\vcenter{\offinterlineskip
  \tableaustep=\tableauside\advance\tableaustep by-\tableaurule
  \kern\normallineskip\hbox
    {\kern\normallineskip\vbox
      {\gettableau#1 0 }%
     \kern\normallineskip\kern\tableaurule}%
  \kern\normallineskip\kern\tableaurule}}
\def\gettableau#1 {\ifnum#1=0\let\next=\null\else
  \squares{#1}\let\next=\gettableau\fi\next}

\newcommand{\bref}[1]{\textbf{\ref{#1}}}

\def\cF{\mathcal{F}}

\def\cL{\mathcal{L}}

\def\cO{\mathcal{O}}

\def\cV{\mathcal{V}}

\numberwithin{equation}{section} \makeatletter
\@addtoreset{equation}{section}

\def\ads{AdS$_{3}\;$}
\def\be{\begin{equation}}
\def\ee{\end{equation}}
\def\ba{\begin{array}}
\def\ea{\end{array}}

\def\dps{\displaystyle}

\def\ba{\begin{array}}
\def\ea{\end{array}}

\def\dps{\displaystyle}

\def\calpha{\alpha}
\def\cbeta{\beta}
\def\tdelta{\Delta_{p}}

\def\etdelta{\epsilon_p}
\def\rads{R}
\def\length{\cL_{_{\hspace{-0.5mm}AdS}}}
\def\lengthA{\cL_{_{\hspace{-0.5mm}AdS_{_3}[2]}}}
\def\lengthB{\cL_{_{\hspace{-0.5mm}AdS_{_3}[3]}}}
\def\lengthG{\cL_{_{\hspace{-0.5mm}AdS_{_3}[n-k]}}}
\def\banados{the  Ba\~{n}ados metric\;}
\def\mobius{M\"obius\;}

\usepackage{jheppub}
\makeatletter
\def\@fpheader{\vspace{-.1cm}}
\makeatother

\title{Four-point conformal blocks with three heavy background operators}

\author[a,b]{Konstantin\ Alkalaev}

\author[a]{Mikhail\ Pavlov}

\affiliation[a]{I.E. Tamm Department of Theoretical Physics, \\P.N. Lebedev Physical
Institute,\\ Leninsky ave. 53, 119991 Moscow, Russia}
\affiliation[b]{Department of General and Applied Physics, \\
Moscow Institute of Physics and Technology, \\
Institutskiy per. 7, Dolgoprudnyi, \\141700 Moscow region, Russia}
\emailAdd{alkalaev@lpi.ru}
\emailAdd{pavlov@lpi.ru}

\abstract{We study  CFT$_2$ Virasoro conformal blocks of the 4-point correlation function $\langle \cO_L \cO_H \cO_H \cO_H   \rangle $ with
three background  operators $\cO_H$ and one perturbative operator $\cO_L$ of  dimensions   $\Delta_L/\Delta_H \ll1$.  The conformal block
function is calculated in the large central charge limit using the monodromy method. From the holographic perspective, the background
operators create $AdS_3$ space with three conical singularities parameterized by dimensions $\Delta_H$, while the perturbative operator
corresponds to the geodesic line stretched from the boundary to the bulk. The geodesic length calculates the perturbative conformal block. We
propose how to address the block/length correspondence problem in  the general case of higher-point correlation functions $\langle \cO_L
\cdots \cO_L \cO_H \cdots \cO_H   \rangle $  with arbitrary numbers of background and perturbative operators.

}

\makeatletter
\def\@fpheader{\vspace{-.1cm}}
\makeatother

\begin{document}

\maketitle
\flushbottom

\section{Introduction}

The AdS/CFT correspondence for large-$c$ conformal blocks in $CFT_2$ was  originally studied  in  the framework of the heavy-light
approximation when two of primary operators produce the conical singularity/BTZ metric in the bulk \cite{Asplund:2014coa,Fitzpatrick:2014vua}.
The other operators are considered as perturbations and can be realized as massive  particles propagating on the background space
\cite{Hijano:2015rla,Fitzpatrick:2015zha,Hijano:2015qja,Alkalaev:2015wia,Banerjee:2016qca,Chen:2016dfb,Alkalaev:2016rjl,Beccaria:2015shq,deBoer:2014sna,Hulik:2016ifr},
or, more geometrically, as the weighted Steiner trees in hyperbolic geometry \cite{Alkalaev:2018nik}.

In this paper,  the case of more than two background operators is considered. We take  the $s$-channel conformal block of the 4-point
correlation function with three background operators and one perturbative operator,
\be
\label{LHHH}
\langle \cO_L(z,\bar{z}) \cO_H(0)  \cO_H(1) \cO_H(\infty) \rangle\;,
\ee
where $(z,\bar z)\in \mathbb{C}$, and the conformal dimensions are such that
\be
\label{lightness1}
\frac{\Delta_{L,H}}{c} =\text{fixed at}\;\;c\to \infty\;\;\;\text{and} \;\;\;\frac{\Delta_{L}}{\Delta_{H}} \ll 1\;.
\ee
Formulating the heavy-light  approximation with  three background operators  and using the monodromy method we explicitly calculate the
large-$c$ 4-point conformal block in the first order in the lightness parameter $\Delta_{L}/\Delta_H$. The zeroth order is given by the
3-point function of the background operators $\cO_H$ that create the bulk geometry identified with   AdS$_3$ space with three conical defects.
\vspace{-5mm}
\begin{figure}[H]
  \centering
  \begin{minipage}[h]{0.35\linewidth}
    \includegraphics[width=1\linewidth]{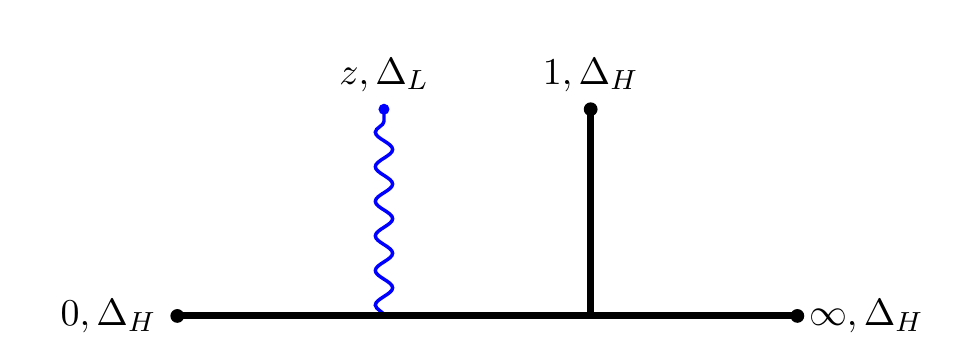}
  \end{minipage}
  \qquad\qquad
   \centering
  \begin{minipage}[h]{0.15\linewidth}
    \includegraphics[width=1\linewidth]{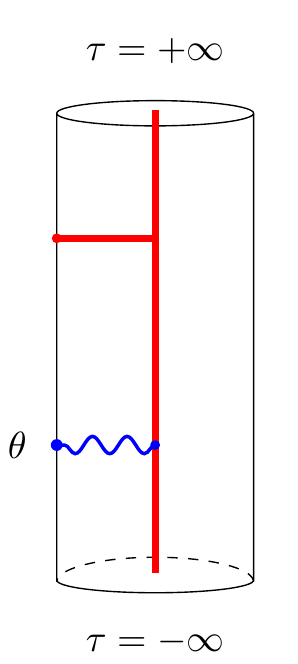}
  \end{minipage}
\caption{The 4-point HHHL block and its holographically dual realization in the three dimensional bulk (a rigid cylinder). The red lines
inside the cylinder visualize  the 3-point function $\langle \cO_H \cO_H \cO_H\rangle$ of heavy operators that created this conical defect
geometry. The wavy blue line denotes the perturbative operator $\cO_L$ propagating in the background. }
\label{figure}
\end{figure}

The operator $\cO_L(z,\bar z)$ is represented as the geodesic line stretched from the conformal boundary to a distinguished point in the bulk,
see Fig. \bref{figure}. Then, the geodesic length calculates the large-$c$ conformal block of the HHHL correlation function \eqref{LHHH} in
the first order of the perturbative expansion  \eqref{lightness1}.

The paper is organized as follows. In Section \bref{sec:monodromy} we calculate the large-$c$ 4-point block within the heavy-light
perturbation theory using the monodromy method. In Section \bref{sec:maps}, based on  \banados of locally AdS$_3$ spaces, we discuss different
coordinate systems in the bulk induced by the boundary conformal (Schwarz) maps.  Section \bref{sec:3pt}  explicitly re-considers the known
case of the 3-point function with two heavy insertions and shows that the block is calculated by the geodesic length of the boundary-to-bulk
line. Section \bref{sec:4pt} follows the same patter and shows that the geodesic segment in the geometry created by three heavy insertions
calculates the 4-point  perturbative block found in Section \bref{sec:block}. In the concluding Section \bref{sec:conclusion} we propose how
to generalize the obtained results to higher-point conformal blocks with more than three background operators.

\section{Perturbative large-$c$ conformal blocks}
\label{sec:monodromy}

Let us consider  primary operators $\cO_i$ in the plane CFT$_2$ with conformal dimensions $(\Delta_i, \bar \Delta_i)$, $i=1,...,4$. The
4-point correlation function in the $s$-channel can be expanded into conformal blocks   \cite{Belavin:1984vu}
\be
\label{bfc}
\langle \cO_{1}(0)\cO_2(z,\bar z)\cO_3(1) \cO_4(\infty)\rangle  = \sum_p C_{12p} C_{34p}\, \cF_{p}(z) \bar{\cF}_{p}(\bar{z})\;,
\ee
where $C_{ijp}$ are structure constants, the (holomorphic) conformal blocks $\cF_{p}(z) = \cF(z|\Delta_i, \tdelta, c)$ in a given channel $p$
depend on external $\Delta_i$, $i=1,...,4$ and intermediate  $\tdelta$ conformal dimensions, and on the central charge $c$, see Appendix
\bref{app:A}.

We consider the large central charge $c \to \infty$. Suppose that all the primary operators are heavy, i.e. their conformal dimensions grow
linearly with the charge, $\Delta_i/c$ are fixed. Then, the 4-point block is exponentiated as
\be
\label{class}
\cF(z|\Delta_i, \tdelta, c)\; \approx \;  \exp{\frac{c}{6} f(z|\epsilon_i, \etdelta)}
\quad \text{at}\quad c\to \infty\;,
\ee
where $f(z|\epsilon_i, \etdelta)$ is the {\it classical} conformal block depending on external and intermediate  classical dimensions
$\epsilon_i = 6\Delta_i/c$ and $\etdelta = 6\tdelta/c$ \cite{Zamolodchikov1986}.

The large-$c$ conformal blocks \eqref{class} can be calculated  using the monodromy method.\footnote{For review and recent studies of the
monodromy method  see e.g. \cite{Harlow:2011ny,Hartman:2013mia,Fitzpatrick:2014vua,Alkalaev:2015lca,Alkalaev:2016rjl,Kusuki:2018nms}.}  To
this end, one considers the BPZ equation for the 5-point correlation function
\be
\label{5pt_cor}
\langle \cO_{1}(0,0)\cO_2(z,\bar z)\Psi(y,\bar y)\cO_3(1,1) \cO_4(\infty,\infty)\rangle\;,
\ee
obtained from the original function \eqref{bfc} by inserting the degenerate operator $\Psi(y,\bar y)$ of conformal  dimension $\Delta_{(1,2)}
= -\frac{1}{2} - \frac{9}{2c}$ in some point $(y,\bar y)\in \mathbb{C}$ \cite{Belavin:1984vu}. Contrary to the original heavy operators this
new operator is light because $\Delta_{(1,2)} = \cO(c^0)$ at $c\to\infty$.  Now, the 5-point correlation function \eqref{5pt_cor}
can be expanded into  conformal  blocks in the OPE channel when the degenerate operator is inserted between two intermediate channels. The
resulting  5-point conformal block is $\cV(y,z|\Delta_i,\Delta_{(1,2)},\tdelta,\Delta_k,c)$, where the intermediate dimensions are related by
the fusion relation $\tdelta- \Delta_k = - \frac{9}{2c}\pm \frac{1}{2}\sqrt{1-\frac{24 \tdelta }{c}}$.

In the large-$c$ limit, the fusion rules claim that the 5-point block factorizes into the original 4-point block $\cF(z|\Delta_i, \tdelta, c)$
and a bi-local prefactor  $\psi(z,y)$ as
\be
\label{5block}
\cV(y,z|\Delta_i,\Delta_{(1,2)},\tdelta,c) \; \approx \;   \psi(y,z) \exp\left[\frac{c}{6} f(z|\epsilon_{i},\etdelta)\right] \quad
\text{at}\quad c\to \infty \;,
\ee
where we have taken into account that the 4-point block can be  exponentiated \eqref{class}.

From the BPZ equation for the original 5-point conformal block \eqref{5block}  one finds that the prefactor $\psi(y,z)$ satisfies the Fuchsian
equation
\be
\label{fe}
\dps \left[\frac{d^2}{dy^2} + T(y,z)\right] \psi(y,z)=0 \;,
\ee
with the energy-momentum tensor of the form
\be
\label{T}
\dps T(y,z) = \frac{\epsilon_1}{y^2} + \frac{\epsilon_2}{(y-z)^2}+ \frac{\epsilon_3}{(1-y)^2}  + \frac{\epsilon_4-\epsilon_3 - \epsilon_1 -
\epsilon_2}{(y-1)y}+ c_2\,\frac{(1-z)z}{y(1-y)(y-z)}\;, \ee
where the {\it accessory parameter} is expressed by
\be
\label{parameter}
c_2 = \frac{d}{dz} f(z|\epsilon_{i},\etdelta)\;.
\ee
Note that there are three more accessory parameters associated to the point $0,1,\infty$ along with the energy-momentum tensor explicitly
depending on them. However, recalling that $T \sim z^{-4}$ at infinity we can isolate the only independent accessory parameter $c_2$ so that
the resulting $T$ is given by \eqref{T}.

The monodromy method considers the Fuchsian equation \eqref{fe} with {\it a priori} independent accessory parameter having no link to the
4-point conformal block. Comparing monodromy matrices of the original 5-point block and solutions to the Fuchsian equation yields the
algebraic equation on the accessory parameter that expresses $c_2$ as a function of coordinates and conformal dimensions. Then, recalling the
relation  \eqref{parameter} we can explicitly integrate to obtain the 4-point large-$c$ conformal block.

\subsection{Perturbative expansion}

Suppose that the conformal dimensions are organized as follows
\be
\label{lightness2}
\Delta_2/\Delta_{1,3,4} \ll1 \quad \text{and} \quad \Delta_1 \sim \Delta_3 \sim \Delta_4\;,
\ee
i.e. there are three background operators with dimensions of the same order and one perturbative operator. In this way we obtain the
correlation function of the type \eqref{LHHH} -- \eqref{lightness1}.

Assuming  \eqref{lightness2} the Fuchsian equation \eqref{fe} - \eqref{parameter}  can be explicitly solved by  expanding all functions up to
the first order in $\Delta_2$ as
\be
\label{decos}
\begin{gathered}
\psi(y,z) = \psi^{(0)}(y,z) + \psi^{(1)}(y,z)+...\,,
\qquad
T(y,z) = T^{(0)}(y,z) + T^{(1)}(y,z)+...\,,
\\
f(z|\epsilon, \etdelta) = f^{(0)}(z|\epsilon, \etdelta) + f^{(1)}(z|\epsilon, \etdelta)  + ...\,,
\qquad
c_2(z|\epsilon, \etdelta) = c_2^{(0)}(z|\epsilon, \etdelta) + c_2^{(1)}(z|\epsilon, \etdelta)  + ...\,.
 \end{gathered}
\ee

A few comments are in order. The term $f^{(0)} =0$ because the conformal block for the 3-point function $\langle \cO_H(0) \cO_H(1)
\cO_H(\infty)\rangle $ is equal to 1 that directly follows from  \eqref{3ptcl}.  The zeroth order  accessory parameter is also zero,
$c_2^{(0)}(z|\epsilon, \etdelta)=0$. Since we will consider only first-order corrections then the notations can be simplified by denoting
$c_2^{(1)}(z|\epsilon, \etdelta) \equiv  c_2(z|\epsilon, \etdelta)$.

The Fuchsian equation in the lowest orders takes the form\footnote{Note that the heavy-light perturbation expansion used to solve the Fuchsian
equation is quite natural but not the only possible perturbation approach. The other perturbation expansion scheme can be found in
\cite{Hadasz:2006rb}. See also the study of different limits of 4-point heavy-light conformal blocks \cite{Kusuki:2018wcv}.}
\begin{align}
\label{zeroth}
\left[\frac{d^2}{dy^2} + T^{(0)}(y)\right]\psi^{(0)}(y,z) &= 0\;,\\
\label{first}
\left[\frac{d^2}{dy^2} + T^{(0)}(y)\right]\psi^{(1)}(y,z) &= - T^{(1)}(y,z) \psi^{(0)}(y,z)\;,
\end{align}
where the zeroth-order energy-momentum tensor $T^{(0)}(y)$ and the first-order correction $T^{(1)}(y)$ are given by
\begin{align}
\label{pe0}
T^{(0)}(y) & = \frac{\epsilon_1}{y^2} + \frac{\epsilon_3}{(1-y)^2} + \frac{\epsilon_1 +\epsilon_3-\epsilon_4}{y(1-y)}\;,\\
T^{(1)}(y,z)& = c_2\,\frac{(1-z)z}{y(1-y)(y-z)} + \frac{\epsilon_2}{(y-z)^2}+\frac{\epsilon_2}{y(1-y)}\;.
\end{align}
Note that $T^{(1)}(y,z)$  is indeed the first order correction  because $c_2 = \cO(\epsilon_2)$.

\subsection{First-order solution}

The zeroth-order Fuchsian equation \eqref{zeroth} can be reduced to the hypergeometric equation  in the so-called  Q-form \cite{Q-form}. The
parameters of the resulting hypergeometric function are expressed in terms of the conformal dimensions $\epsilon_{i}$ and $\etdelta$.  To
simplify the further analysis  we assume that
\be
\Delta_3 = \Delta_4\;.
\ee
Moreover, this condition is required when considering going from three to two background operators. There are two branches of the zeroth-order
solution,
\be
\label{zo}
\dps \psi^{(0)}_{\pm}(y) = (1-y)^{\frac{1+\alpha}{2}} y^{\frac{1\pm\cbeta}{2}} F_{\pm}(\alpha,\beta|y) \,,
\ee
where the hypergeometric functions are given by
\be
\label{Fpm}
F_{\pm}(\alpha,\beta|y) ={}_2F_1\left(\frac{1\pm\cbeta}{2},\frac{1\pm\cbeta}{2}+ \alpha, 1\pm\cbeta,y\right)\,,
\ee
and
\be
\label{ab}
\calpha =\sqrt{1-\frac{24\Delta_4}{c}}\;,\qquad \cbeta =\sqrt{1-\frac{24\Delta_1}{c}}\;,
\qquad \;\;
0<\alpha,\beta< 1\;.
\ee

Consider then the Fuchsian equation in the first order \eqref{first}. Using the method of variation of parameters we find the first order
correction,
\be
\label{fo}
\dps \psi^{(1)}_{\pm}(y,z) = \psi^{(0)}_{+}(y)\int dy \frac{\psi^{(0)}_{-} T^{(1)}(y,z)\psi^{(0)}_{\pm}}{W} - \psi^{(0)}_{-}(y)\int dy
\frac{\psi^{(0)}_{+} T^{(1)}(y,z)\psi^{(0)}_{\pm}}{W}\;,
\ee
where the Wronskian is given by
\be
\label{wronskian}
W \equiv - \psi^{(0)}_{+} (y) \frac{d \psi^{(0)}_{-}}{dy} + \psi^{(0)}_{-} (y) \frac{d \psi^{(0)}_{+}}{dy}  = \frac{\sin \pi \cbeta}{\pi} \;.
\ee

Thus, the first-order solution  reads as $\psi_{\pm}(y,z) = \psi_{\pm}^{(0)}(y,z) +\psi_{\pm}^{(1)}(y,z)$, where $\psi_{\pm}^0$ and
$\psi_{\pm}^1$ are given by \eqref{zo} and \eqref{fo}. It is parameterized by the background dimensions (through $\alpha,\beta$) and depends
on the indeterminate accessory parameter $c_2$.

\subsection{More on the zeroth-order solution}
\label{sec:zeroth}

When solving the second-order Fuchsian equation we can equally choose a linear combination of the  solutions $\psi_{\pm}$. In particular, in
the zeroth order,
\be
\label{basis}
\psi^{(0)}_{i}(y) \rightarrow  \hat\psi^{(0)}_{i}(y)  = A_{ij} \psi^{(0)}_{j}(y)\;, \qquad
\ee
where $i,j = +,-$ and $A_{ij}$ is a non-degenerate $[2\times 2]$ matrix. Let
$A_{ij} = \begin{pmatrix}
                                                                                 \alpha & 1 \\
                                                                                 0 & 1
                                                                               \end{pmatrix}$. Expanding the resulting solution in powers of
                                                                               $\epsilon_1$ we find that
\be
\label{3bas}
\hat\psi^{(0)}_{\pm}(y) = (1-y)^{\frac{1 \pm \alpha}{2}} + \cO(\epsilon_1)\;.
\ee
In this way, we reproduce the case of two background operators \cite{Fitzpatrick:2014vua} obtained here by sending $\epsilon_1 \to 0$. Indeed,
operators $\cO_{1,2}$ in the 4-point correlation function \eqref{bfc} are now perturbatively light compared to $\cO_{3,4}$ and the heavy-light
perturbative expansion assumes that in the zeroth order $\epsilon_1 = \epsilon_2 = 0$. Then, having started with the original $5$-point
correlation function \eqref{5pt_cor} now we are left with the 3-point function of the form
\be
\label{BPZ_3pt}
\langle \Psi_{(1,2)}(y,\bar y)\cO_H(1) \cO_H(\infty)\rangle\;.
\ee
The BPZ equation for \eqref{BPZ_3pt} in the large-$c$ limit is just the Fuchsian equation \eqref{zeroth}, \eqref{pe0} taken at
$\epsilon_3=\epsilon_4$ and $\epsilon_1 = 0$. On the other hand, the BPZ equation is solved by functions $(1-y)^{\frac{1 \pm \alpha}{2}}$ (see
e.g. \cite{DiFrancesco:1997nk}). Thus, the  functions $\psi^{(0)}$ are identified with 3-point degenerate conformal block.

Let us now consider the present  case of three background operators. In the zeroth order, when  $\Delta_2=0$,   we obtain from \eqref{5pt_cor}
the 4-point function
\be
\label{BPZ_function}
\langle \Psi_{(1,2)}(y,\bar y)\cO_H(0) \cO_H(1) \cO_H(\infty)\rangle\;.
\ee
Since the operator $\Psi_{(1,2)}$ is degenerate (on the second level) then the associated BPZ equation for the degenerate 4-point conformal
block in the $s$-channel  reads \cite{Belavin:1984vu}
\begin{multline}
\label{BPZ}
\left[\frac{3}{4\Delta_{(1,2)}+2}\frac{d^2}{dy^2} + \frac{2-4y}{y(1-y)}\frac{d}{dy} + \frac{\Delta_1 + \Delta_3 -
\Delta_4+\Delta_{(1,2)}}{y(y-1)}-\frac{\Delta_1}{y^2} - \frac{\Delta_3}{(1-y)^2}\right]\cF(y|\Delta_{1,3,4},\tdelta,c) = 0\;.
\end{multline}
It is well known that the solution is given by the hypergeometric functions.
In the large-$c$ regime we rescale the dimensions and notice then that the second term in \eqref{BPZ} is  small compared to the others. The
resulting equation takes the form
\be
\left[\frac{d^2}{dy^2} + \frac{\epsilon_1}{y^2} + \frac{\epsilon_3}{(1-y)^2} + \frac{\epsilon_3 + \epsilon_1 - \epsilon_4}{y(1-y)}\right]
\cF(y|\epsilon_{1,3,4},\etdelta) =0 \;.
\ee
This equation coincides with the zeroth-order Fuchsian equation \eqref{zeroth}, \eqref{pe0}. Thus, the zeroth-order solution $\psi^{(0)}$  is
the $4$-point degenerate  conformal block.

\subsection{Accessory parameters and the conformal block}
\label{sec:block}
Let us consider a contour $\Gamma$ enclosing points $0$ and $z$,  and calculate the corresponding monodromy  of the first order solution
$\psi_{\pm}(y,z) = \psi_{\pm}^{(0)}(y,z) +\psi_{\pm}^{(1)}(y,z)$. The monodromy matrix $M=||M_{ij}||$ is $[2\times 2]$ matrix, $i = (+,-)$,
that can be decomposed as
\be
\label{fb}
M_{ij} = M_{ij}^{(0)} + M_{ij}^{(1)} +...\;,
\ee
where the zeroth and the first order terms are given by
\be
M^{(0)} = - \begin{pmatrix}
  e^{i \pi  \cbeta }& 0\\
  0& e^{-i \pi   \cbeta}
\end{pmatrix} \,,\qquad
M^{(1)} = - \begin{pmatrix}
  e^{i \pi  \cbeta } I_{++}\;\;& e^{-i \pi  \cbeta }I_{+-}\\
  e^{i \pi  \cbeta } I_{-+}\;\;&  e^{-  i \pi  \cbeta }I_{--}
\end{pmatrix} \,,
\ee
where
\be
I_{+\pm} =  \frac{\pi}{\sin(\pi \cbeta)}\int_{\Gamma} dy \; \psi^{(0)}_{-} T^{(1)}(y,z) \psi^{(0)}_{\pm}\;,
\qquad
I_{-\pm} = -\frac{\pi}{\sin(\pi \cbeta)}\int_{\Gamma} dy \; \psi^{(0)}_{\pm} T^{(1)}(y,z) \psi^{(0)}_{-}\;.
\ee
Denoting   $X = z(c_2(1-z) - 2\epsilon_2) + \epsilon_2(1- \alpha z)$ and
$Y = X- \epsilon_2 \cbeta (1-z)$, the contour integrals can be represented in the form
\be
\label{I}
\ba{l}
\hspace{7mm}\dps I_{++} =\frac{2 i \pi^2 (1-z)^{\alpha} F_{+} F_{-}}{\sin \pi \cbeta} \left( X + \epsilon_2 z (1 - z)\frac{d \log(F_{+}
F_{-})}{dz} \right)\;,
\\
\\
\dps I_{+-} = - \frac{2i \pi^2 (1-z)^{\alpha} z^{-\cbeta}F_{-}F_{-}}{\sin\pi \cbeta}\left( Y  +  \epsilon_2 z (1 - z)\frac{d \log
(F_{-}F_{-})}{dz}\right) \;,
\ea
\ee
and
\be
I_{--} = - I_{++}\;,
\qquad
I_{-+} = I_{+-}\big|_{\cbeta\to -\cbeta}\;,
\ee
where $F_{\pm}$ are given in \eqref{Fpm}.

On the other hand, traversing the degenerate operator $\Psi_{(1,2)}(y,\bar y)$ along the contour $\Gamma$, i.e. around the primary operators
$\cO(0)$ and $\cO(z,\bar z)$ we find that the respective monodromy is diagonal. Since the $5$-point  block \eqref{5block} has two independent
components, the monodromy matrix  $\widetilde M=||\widetilde M_{ij}||$  is also $[2\times 2]$, $i = (+,-)$. Taking the large-$c$ limit and
calculating the matrix  along the contour $\Gamma$ we find an exact expression
\be
\label{TM}
\widetilde{M} = - \begin{pmatrix}
  e^{i \pi  \gamma }& 0\\
  0& e^{-i \pi   \gamma}
\end{pmatrix},
\qquad \;\;
\gamma = \sqrt{1-\frac{24\tdelta}{c}}\;,
\qquad
0<\gamma< 1\;,
\ee
which is valid in any order of our perturbation theory.

The two monodromy matrices $M$ \eqref{fb}  and $\widetilde M$ \eqref{TM} describe the same monodromy and, therefore, have equal eigenvalues.
The matrix $\widetilde M$ is already diagonal.  To diagonalize $M$ we consider the corresponding  characteristic equation
\be
(\exp[i \pi \cbeta] (1 + I_{++}) - \lambda)(\exp[- i \pi \cbeta] (1 - I_{++}) - \lambda) -  I_{+-}I_{-+} =0\;,
\ee
which has two complex conjugated roots
\be
\lambda_{\pm} = -  \frac{\exp[i \pi \cbeta] (1 + I_{++}) +   \exp[-i \pi \cbeta] (1 - I_{++}) \pm i \sqrt{-D}}{2}\;,
\ee
with discriminant
\be
D = (\exp[i \pi \cbeta] (1 + I_{++}) +   \exp[-i \pi \cbeta] (1 - I_{++}))^2 - 4 + 4(I^2_{++} + I_{+-}I_{-+})\;.
\ee
In order to  compare the eigenvalues $\lambda_{\pm}$ with those of \eqref{TM} one expands in powers of $\epsilon_2$. Since the monodromy
integrals are linear in $\epsilon_2$ (recall that $c_2$ in $I_{++}$ is the first order correction, see our comment below \eqref{decos}) we get
in the first order,
\be
\begin{gathered}
\lambda_{+} = - e^{i \pi \cbeta}\left(1 + I_{++}\right) +\cO(\epsilon_2^2)\;,\\[5pt]
\lambda_{-} = - e^{-i \pi \cbeta}\left(1 - I_{++}\right) +\cO(\epsilon_2^2)\;,
\end{gathered}
\ee
that must be equal to $-e^{\pm i \pi \gamma}$. Equating order by order we obtain the following constraints
\be
\label{ae}
\gamma = \beta\;,
\qquad
I_{++}  =  0\;,
\ee
where $I_{++}$ is given in \eqref{I}. From the first relation in \eqref{ae} we find out that the first and the intermediate conformal
dimensions must be equal,
\be
\label{fus}
\Delta_p = \Delta_1\;.
\ee
Such a condition is to be expected within the heavy-light perturbative expansion. By the fusion rules, when $\Delta_2 =0$, the intermediate
channel should be equated to the first primary insertion, thereby losing its own character just because the resulting 3-point function has no
exchange channels.

Now, solving the second relation in \eqref{ae} we obtain
\be
c_2 = \epsilon_2 \left[ \frac{1+\alpha}{1-z} - \frac{1}{z} - \frac{d \log(F_{+} F_{-})}{dz}\right]\;.
\ee
Integrating the defining relation \eqref{parameter} we  find the 4-point perturbative classical conformal block with three background
insertions
\be
\label{thfb}
f(z|\alpha, \beta,\epsilon_2)= - \epsilon_2\Big(\log(1-z)^{1+\alpha} + \log z + \log F_{+}(\alpha,\beta|z)+ \log F_{-}(\alpha,\beta|z)
\Big)\;,
\ee
where functions $F_{\pm}$ are given by \eqref{Fpm}.

A few comments are in order. First, in the limit $\beta \to 1 $ corresponding to $\Delta_1\to 0$ we reproduce the 4-point perturbative block
with two background operators at $\epsilon_1 = \epsilon_p$ \cite{Hijano:2015rla}.  To this end, we decompose the function \eqref{thfb}, where
$\beta = \beta(\epsilon_1)$ is given by \eqref{ab}, in powers of $\epsilon_1$ and keep the linear terms (see also Appendix \bref{app:A}).
Secondly, we note that when calculating the monodromy matrix $M$ we could equally use solutions \eqref{basis}. In this case, the resulting
monodromy matrix is related to \eqref{fb} by $ \hat M = A M A^{-1}$ and, therefore, the eigenvalue problem remains the same giving rise to the
same accessory parameter and conformal block.

\section{Holographically  dual description}
\label{sec:bulk}

In this section we develop the holographic interpretation of the large-$c$ perturbative conformal blocks with three background operators. For
two background operators the respective blocks were realized as geodesic trees in the conical defect space parameterized by $\alpha =
\alpha(\Delta_H)$ \eqref{ab}.

\subsection{Conformal maps and the bulk metrics}
\label{sec:maps}
In the AdS$_3$/CFT$_2$ correspondence, the locally AdS$_3$ geometry created by heavy insertions of the boundary CFT can be described in the
Ba\~{n}ados form  \cite{Banados:1998gg}
\be
\label{Banados}
ds^2 = \rads^2\left( -H dz^2 -\bar{H}d\bar{z}^2 + \frac{u^2}{4}\,  H \bar{H} \, dz d\bar z+ \frac{du^2 + dz d\bar{z}}{u^2} \right)\;,
\ee
with $u\in [0,\infty)$ and $z, \bar z \in \mathbb{C}$ being local coordinates, the radius is $\rads$. Arbitrary (anti)ho\-lo\-morphic
functions $H=H(z)$ and $\bar H =\bar H(\bar z)$ can be interpreted as components of the holographic CFT$_2$ energy-momentum  tensor
\be
\label{TH}
T(z) = \frac{c}{6} H(z)\;,
\ee
where the central charge is $c = 3\rads/2G_N$ \cite{Brown:1986nw,Balasubramanian:1999re}. Under $z\to w(z)$ it transforms in the standard
fashion as
\be
\label{trans}
T(z) = \left(w^\prime\right)^2 T(w) + \frac{c}{12}\, \{w,z\}\;,
\qquad
\text{where}
\quad
\{w,z\} =  \frac{w'''}{w'}  - \frac{3}{2}\left(\frac{w''}{w'} \right)^{2}\;,
\ee
where the prime denotes differentiation with respect to $z$.  The anti-holomorphic component transforms analogously.

Let us find a map $z \to w(z)$ such that $H(w(z)) = 0$. Away from singularities it would correspond to pure AdS$_3$ in the Poincare
coordinates, cf.  \eqref{Banados}.  This can be achieved provided that
\be
\label{def_eq}
H(z)  = \half \{w,z\}\;.
\ee
The solution to the above equation  can be represented  as the ratio of two independent solutions to the auxiliary Fuchsian equation
$\psi^{''}+H \psi =0$ (see e.g. \cite{Q-form}). This is the so-called Schwarz map
\be
\label{wz}
w(z)= \frac{A\, \psi_1(z) + B \,\psi_2(z)}{C\, \psi_1(z) + D\, \psi_2(z)}
\equiv
\frac{A\, \frac{\psi_1(z)}{\psi_2(z)} + B }{C\, \frac{\psi_1(z)}{\psi_2(z)} + D} \;, \qquad A D  - B C \neq 0\;,
\ee
where $\psi_{1,2}$ are two independent Fuchsian solutions, and  $A,B,C,D\in \mathbb{C}$ parameterize the M\"obius transformation of
$\psi_1(z)/\psi_2(z)$.

The  relation  \eqref{TH} and the Fuchsian interpretation of solutions to the equation \eqref{def_eq} suggest  that in the large-$c$ regime
the function $H$ can be identified with the {\it classical} energy-momentum tensor arising  in the Fuchsian equation \eqref{zeroth} of the
monodromy method, i.e.,
\be
\label{HT0}
H(z|{\bf z}) \equiv T^{(0)}(z|{\bf z})\;,
\ee
where we introduced  the set of singular points ${\bf z}$ identified with locations of the background operators.

Since any solution to the Einstein equations with the (negative) cosmological constant is locally AdS$_3$ space then the boundary map
\eqref{wz} can be extended to the whole three-dimensional space, $w = w(z,\bar z,u)$, $\bar w = \bar w(z,\bar z,u)$,  and $v = v(z, \bar
z,u)$, such that the resulting metric describes the Poincare patch
\be
\label{PP}
d\tilde{s}^2  = \frac{dv^2 + dw d\bar{w}}{v^2}\;.
\ee
The explicit coordinate transformation reads \cite{Roberts:2012aq}
\be
\label{roberts}
\begin{gathered}
\dps w(z,\bar z,u) = w(z)  - \frac{2 u^2 w^\prime(z)^2 \bar w^{\prime\prime}(\bar z)}{4w^\prime(z)\bar w^{\prime}(\bar z)+u^2
w^{\prime\prime}(z)\bar w^{\prime\prime}(\bar z)} \;,
\\[5pt]
\dps \bar w(z,\bar z,u) = \bar w(\bar z)  - \frac{2 u^2 \bar w^\prime(\bar z)^2 w^{\prime\prime}(z)}{4w^\prime(z)\bar w^{\prime}(\bar z)+u^2
w^{\prime\prime}(z)\bar w^{\prime\prime}(\bar z)} \;,
\\[5pt]
\dps v(z,\bar z,u) =  u\, \frac{4\left( w^\prime(z)\bar w^\prime(\bar z)\right)^{3/2}}{4w^\prime(z)\bar w^{\prime}(\bar z)+u^2
w^{\prime\prime}(z)\bar w^{\prime\prime}(\bar z)}\;.
\end{gathered}
\ee

The length of a geodesic line stretched between two points $(w_1, \bar w_1, v_1)$ and $(w_2, \bar w_2, v_2)$  evaluated in the Poincare
coordinates \eqref{PP} is particularly simple
\be
\label{geodesic}
\length = \rads\log\frac{(w_1 - w_2)(\bar w_1 - \bar w_2)}{v_1 v_2}\;.
\ee


Finally, we note that in the Euclidean case the Poincare patch covers the whole global \ads space
\be
\label{GADS}
d\hat {s}^2  = \frac{d\tau^2+ d \rho^2 +\sin^2\rho d\phi^2}{\cos^2 \rho}\;.
\ee
through the coordinate change  $w = e^{\theta} \sin \rho$, $\bar w= e^{\bar \theta} \sin \rho$,
$v = e^{\frac{\theta+\bar\theta}{2}} \cos \rho$,
where $\theta = \tau+i \phi$ and $\rho$ are coordinates of the global AdS$_3$     (rigid cylinder).  The conformal boundary is  at $\rho =
\pi/2$. There is a conformal map $\theta = \log w$ from the boundary $(w,\bar w)$-plane  to the boundary $(\theta, \bar \theta)$-cylinder.


\subsection{3-point HHL block as geodesic length}
\label{sec:3pt}
Let us first consider the simplest case  of 3-point function with two background insertions in $1$ and $\infty$, see \eqref{3}. The classical
energy-momentum tensor \eqref{HT0} is given by
\be
T^{(0)}(z) =  \frac{\epsilon_3}{(1-z)^2}\;,
\ee
where $\epsilon_2 = \epsilon_3$ are classical dimensions of the background operators. It follows that there are two lines of coordinate
singularities in  the Ba\~{n}ados metric \eqref{Banados}: $(z,\bar z, u) = (1,1,u)$ and $(z,\bar z, u) = (\infty,\infty,u)$ at $\forall u\in
\mathbb{R}_+$. The resulting space will be denoted as AdS$_3[2]$.

Modulo \mobius transformations, the conformal mapping \eqref{wz} is given by
\be
\label{CD}
w(z) = (1-z)^\alpha\;,
\ee
where we used the Fuchsian solution \eqref{3bas} and $\alpha$ is given by \eqref{ab}. The function $w(z)$ has  two singular points $1$ and
$\infty$, corresponding to locations of the background operators. It maps the $(z,\bar z)$-plane  onto the $(w,\bar w)$-plane  with an {\it
angle deficit} proportional to $\alpha\in (0,1)$, cf. \eqref{ab}. Near $z=\infty$ we can change $z\to 1/z$ so that  $w(z) \approx
z^{-\alpha}$. Thus, at infinity, there is an {\it angle excess} parameterized by $-\alpha$.

The Poincare coordinates in AdS$_3$[2] can be explicitly read off from \eqref{roberts} (see also \cite{Cresswell:2018mpj})
\begin{align}
&w(z,u) = \; (1-z)^\alpha \;\frac{(1-\alpha^2)u^2 + 4(1-z)(1-\bar z)}{(1-\alpha)^2 u^2 + 4(1-z)(1-\bar z)}\;,  \\[5pt]
&\bar w(\bar z,u) = \; (1-\bar z)^\alpha \;\frac{(1-\alpha^2)u^2 + 4(1-z)(1-\bar z)}{(1-\alpha)^2 u^2 + 4(1-z)(1-\bar z)}\;,  \\[5pt]
&\dps v(z,\bar z,u) =  \;4 \alpha\; \frac{u (1-z)^{\frac{1+\alpha}{2}}(1-\bar z)^{\frac{1+\alpha}{2}}}{(1-\alpha)^2 u^2 + 4(1-z)(1-\bar z)}\;.
\end{align}
Near the singular points the map is approximated by $w = (1- z)^\alpha(1+\cO(1-z))$ along with
\begin{align}
z\to 1\;:\qquad &v(z,\bar z,u) =  \; \frac{4\alpha \,u^{-1}}{(1-\alpha)^2}\; \left[(1-z)(1-\bar
z)\right]^{\frac{\alpha+1}{2}}(1+\cO((1-z)(1-\bar z)))\;,  \\[3pt]
z\to \infty\;:\qquad &v(z,\bar z,u) =  \; \alpha\, u\; (z\bar z)^{\frac{\alpha-1}{2}}(1+\cO(1/(z\bar z)))\;.
\end{align}
Since $\alpha \in (0,1)$ the leading  exponent in the first relation above  is positive, while in the second relation it is negative. Thus,
the lines of coordinate singularities of the Ba\~{n}ados metric are  mapped to two boundary points $(w,\bar w, v) = (0,0,0)$ and $(w,\bar w,
v) = (\infty,\infty,0)$.

Further,  when going to the global \ads space \eqref{GADS} both singularities are mapped to the boundary points $\tau = \pm \infty$. The
coordinate $\phi$ is rescaled by $\alpha$ so that we have a wedge-shaped sector cut from the cylinder and the remaining domain is
$\phi\in[0,2\pi \alpha)$, $\tau\in \mathbb{R}$, $\rho \in \mathbb{R}_+$. This is the standard conical defect in the \ads space.

\vspace{-5mm}
\begin{figure}[H]
  \centering
  \begin{minipage}[h]{0.35\linewidth}
    \includegraphics[width=1\linewidth]{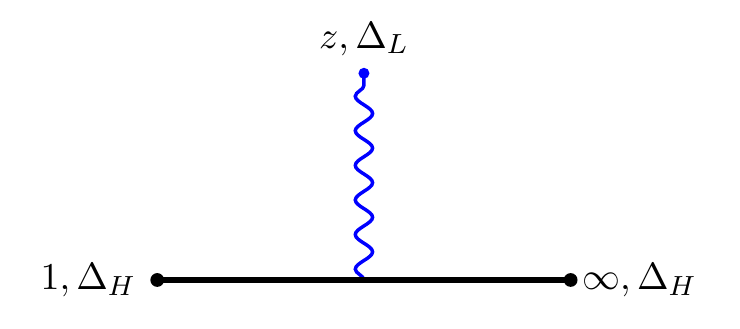}
  \end{minipage}
  \qquad\qquad
   \centering
  \begin{minipage}[h]{0.15\linewidth}
    \includegraphics[width=1\linewidth]{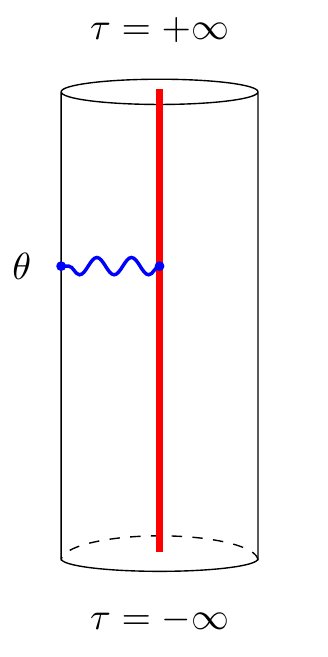}
  \end{minipage}
\caption{The 3-point HHL block and its holographically dual realization in the three dimensional bulk (a rigid cylinder). The red line along
the cylinder axis visualizes  the 2-point function $\langle \cO_H \cO_H\rangle$ of heavy operators inserted  at infinities that produced the
conical defect. The wavy blue line denotes the perturbative operator $\cO_L$ propagating in the background. }
\label{figure2}
\end{figure}

Now we consider the 3-point conformal block, which in the original coordinates is given by
\be
f(z|\epsilon_1) = -\epsilon_1\log(1-z)\;,
\ee
see eq. \eqref{3ptcl}. The perturbative operator of classical dimension $\epsilon_1$ is inserted in $(z,\bar z)$.

To find the block function in different coordinate systems we recall the relation \eqref{class_tr}. In our case, the transformation from some
$x$-coordinates to $y$-coordinates takes the form
\be
f(x) = f(y(x)) + \epsilon_1 \log y'(x)\;,
\ee
where the prime denotes differentiation with respect to $x$.  Keeping only coordinate-dependent terms (this is $\sim$ below) we find that the
block function in three coordinate systems on the boundary  (original $(z, \bar z)$, then conformally mapped $(w, \bar w)$, and global
$(\theta, \bar \theta)$) is given by
\be
\label{3pt_fin}
f(z|\epsilon_1)  \sim  -\epsilon_1\log\frac{w(z)}{w'(z)}\;,
\qquad
f(w|\epsilon_1) \sim -\epsilon_1\log w\;,
\qquad
f(\theta|\epsilon_1) \sim 0\;.
\ee

Let us consider the AdS$_3[2]$ space in the Poincare coordinates and fix two points: the boundary insertion of the perturbative operator
$(w,\bar w, \varepsilon)$ and a distinguished   point inside the bulk $(0,0,1)$, where $\varepsilon\to0 $ is the boundary cut-off. The point
belongs to the line connecting two background insertions at infinities, see Fig. \bref{figure2}.  Then, the geodesic length between these two
points  given by \eqref{geodesic} takes the form
\be
\label{3pt_length}
\lengthA(w,\bar w) = \rads \left(\log w + \log \bar w\right) -\rads\log\varepsilon\;.
\ee
Comparing with \eqref{3pt_fin} we find that in the Poincare coordinates, modulo $\varepsilon$-dependent  terms and constants, the block
function and the geodesic length are identified,
\be
\label{bl3}
f(w|\epsilon_1) \sim -\frac{\;\epsilon_1}{\rads}\; \lengthA(w)\;,
\ee
where we omitted the anti-holomorphic part corresponding to the anti-holomorphic conformal block. In the large-$c$ regime, the standard
formula relating  masses and heavy conformal dimensions takes the form $m R \sim \Delta$ so that the ratio $\epsilon_1/R$ measures the mass of
a point particle propagating in AdS$_3[2]$. This is the block/length identification in the case of 3-point block with two background operators
\cite{Alkalaev:2018nik}.

In the global coordinates, the distinguished point turns to  the axis $\rho=0$ of the rigid cylinder so that the geodesic line is stretched
from the boundary point  to the center, see Fig. \bref{figure2}. Its regulated length turns out to be  zero, $\lengthA(\theta, \bar \theta)
\sim 0$, that agrees with $f(\theta) = 0$ in \eqref{3pt_fin}.

Note that there is another possible geodesic line in the bulk with  boundary endpoints  corresponding to  the CFT operators. It connects two
boundary points and the geodesic length calculates the  4-point identity HHLL conformal block
\cite{Asplund:2014coa,Fitzpatrick:2016mtp,Anand:2017dav}. For conformal blocks with more perturbative insertions the dual geodesic network has
several boundary attachments and one endpoint in the center of the cylinder. In that case both the total length and the perturbative blocks
are non-vanishing satisfying the relation of the type \eqref{bl3}
\cite{Hijano:2015rla,Alkalaev:2015wia,Banerjee:2016qca,Alkalaev:2016rjl,Alkalaev:2018nik}.

\subsection{4-point HHHL block as geodesic length }
\label{sec:4pt}

Let us turn to the case of the background geometry created by three heavy insertions. Here, the classical energy-momentum tensor \eqref{HT0}
is given by
\be
\label{sct}
T^{(0)}(z) =  \frac{\epsilon_1}{z^2} + \frac{\epsilon_3}{(z-1)^2} + \frac{\epsilon_1}{z(1-z)}\;,
\ee
where $\epsilon_1$ and $\epsilon_3 = \epsilon_4$ are classical dimensions of the heavy background insertions in  $0,1, \infty$. The resulting
space defined by the  Ba\~{n}ados metric  \eqref{Banados} will be denoted as AdS$_3[3]$. There are three lines of coordinate singularities:
$(z,\bar z, u) = (0,0,u)$, $(z,\bar z, u) = (1,1,u)$, and  $(z,\bar z, u) = (\infty,\infty,u)$ for any $u\in \mathbb{R}_+$.

Choosing in \eqref{wz} the Fuchsian solutions as $\psi_{1,2}(z)  = \psi_{\pm}^{(0)}(z)$ \eqref{zo} we find that up to the \mobius
transformations  the conformal mapping is given by
\be
\label{schwarz}
w(z) = z^{\beta}\;\frac{_2F_1\left(\frac{1+\beta}{2},\frac{1+\beta}{2}+ \alpha, 1+\beta,z\right)
}{_2F_1\left(\frac{1-\beta}{2},\frac{1-\beta}{2}+ \alpha, 1-\beta,z\right)}\;.
\ee
This is the {\it Schwarz triangle function} that maps the $(z,\bar z)$-plane onto some curvilinear triangle on the $(w,\bar w)$-plane (see
e.g. \cite{Nehari}). Since the \mobius group acts triply transitively and conformally, then the vertices can be placed in any positions, while
the angles remain intact.  By construction, the Schwarz function has three singular points $z=0,1,\infty$ identified with background operator
locations. The angles and verities of the triangle are expressed in terms of parameters $\alpha, \beta$ of the function \eqref{schwarz}. We
find that  the angle in the point $w(0)$ is equal to $\pi\beta$, the second angle in $w(1)$ is equal to $-\pi\alpha$, and the third angle in
$w(\infty)$ is equal to $\pi\alpha$.\footnote{This domain generalizes that one in the case of two singular points, where we had two vertices
with angles $\pm \alpha$, see Section \bref{sec:3pt}.}
Recalling the definition \eqref{ab} we obtain  that the sum of angles in the triangle is $\pi\beta <\pi$.

The Schwarz triangle function near the singular points  can be approximated by\footnote{The most straightforward way to derive these
expansions is to solve the defining equation \eqref{def_eq} using the Frobenius method in the leading orders. Otherwise, one can find
asymptotics of the Schwarz triangle function \eqref{schwarz} by expanding the hypergeometric functions near the singular points.}
\be
\label{asymptotics}
\begin{aligned}
z\to 0: \qquad\;& w(z) \sim   z^{\beta}(1+ \cO(z))\;,\\[2pt]
z\to 1: \qquad\;& w(z) \sim  (1-z)^{-\alpha}(1+ \cO(1-z))\;,\\[2pt]
z\to \infty: \qquad & w(z) \sim  z^{-\alpha}(1+ \cO(1/z))\;, \\[2pt]
\end{aligned}
\ee
where  $\sim$ indicates  that these expansions are valid modulo M\"obius transformations.
The leading  exponents in \eqref{asymptotics} define the angle deficit/excess. The Schwarz triangle function \eqref{schwarz} in the singular
points is given by
\be
\label{values}
w(0) = 0\;,
\qquad
w(1) = \infty\;,
\qquad
w(\infty) = e^{i \pi \beta} \; \frac{\Gamma(1+\beta) \; \Gamma(\frac{1-\beta}{2} + \alpha) \; \Gamma(\frac{1-\beta}{2})}{\Gamma(1-\beta) \;
\Gamma(\frac{1+\beta}{2} + \alpha)\; \Gamma(\frac{1+\beta}{2})}\;.
\ee

From the asymptotics \eqref{asymptotics} we can find how  $v$-coordinate behaves near the singular points. Applying the general relations
\eqref{roberts} we obtain
\begin{align}
z\to 0\;:\qquad &v(z,\bar z,u) =  \; \frac{4\beta \,u^{-1}}{(1-\beta)^2}\; (z\bar z)^{\frac{1+\beta}{2}}(1+\cO(z\bar z))\;,  \\[3pt]
z\to 1\;:\qquad &v(z,\bar z,u) =  \; \frac{4\alpha \,u^{-1}}{(1+\alpha)^2}\; \left[(1-z)(1-\bar
z)\right]^{\frac{1-\alpha}{2}}(1+\cO((1-z)(1-\bar z)))\;,  \\[3pt]
z\to \infty\;:\qquad &v(z,\bar z,u) =  \; \alpha\, u\; (z\bar z)^{-\frac{1+\alpha}{2}}(1+\cO(1/(z\bar z)))\;.
\end{align}
Since $\alpha, \beta \in (0,1)$,  then the leading exponents in the first two relations are positive, while in the third one it is negative.
It follows that the three lines of coordinate singularities in the Ba\~{n}ados metric \eqref{Banados} are mapped to three boundary points in
the Poincare metric which are vertices of the curvilinear triangle. In the global coordinates,\footnote{It would be important to find an
explicit characterization  of the AdS$_3[3]$ in the global coordinates $\theta, \bar \theta, \rho$ by analogy with the conical defect geometry
AdS$_3[2]$. On the other hand, to study $3d$ spaces with conical singularities one recalls that the $3d$ gravity action can be rewritten as
the Liouville field theory on the conformal boundary \cite{Krasnov:2000zq,Krasnov:2000ia}. In particular, a solution with three conical
defects was considered in \cite{Chang:2016ftb}.} they lie  on the boundary cylinder  $\rho=\pi/2$ at $\theta_{0,1} = \pm \infty$ and
$\theta_\infty = \log w_\infty$, where $w_\infty$ is given in \eqref{values}, see Fig. \bref{figure}.

Let us consider now the conformal block \eqref{thfb}  in three boundary coordinate systems: $(z,\bar z)$-plane, $(w,\bar w)$-domain, $(\theta,
\bar \theta)$-cylinder discussed in Section \bref{sec:maps}.   Assuming that we do some coordinate change $x \to x(y)$ the transformation
formula \eqref{class_tr} takes the form
\be
f(x|\alpha, \beta,\epsilon_2) = f(y(x)|\alpha, \beta,\epsilon_2) + \epsilon_2 \log y'(x)\;.
\ee
Differentiating the Schwarz triangle function \eqref{schwarz} and using the Wronskian \eqref{wronskian} one finds that
\be
w^\prime(z) = \frac{\sin\pi \beta}{\pi }\,\frac{(1-z)^{-1-\alpha} z^{-1+\beta}}{F^2_{-}(\alpha,\beta|z)}\;,
\ee
where $F_-$ is given by \eqref{Fpm}. The block function \eqref{thfb} in three coordinate systems is given by
\be
\label{4pt_fin}
f(z|\alpha, \beta,\epsilon_2) \sim -\epsilon_2\log\frac{w(z)}{w'(z)}\;,
\qquad
f(w|\alpha, \beta,\epsilon_2) \sim - \epsilon_2\log w\;,
\qquad
f(\theta|\alpha, \beta,\epsilon_2) \sim 0\;,
\ee
where $\sim$ stands for constants terms, cf. \eqref{3pt_fin}.

The above relations suggest that the bulk interpretation of the 4-point block with three background operators is quite similar to that of
3-point block with two background operators. Indeed,  according to our prescription we fix two points in AdS$_3[3]$ in the Poincare
coordinates: the boundary insertion of the perturbative operator $(w,\bar w, \varepsilon)$ and the distinguished  point in the bulk $(0,0,1)$,
where the cut-off $\varepsilon\to0 $.  The distinguished point belongs to the vertex joining the background heavy insertions: two  at
infinities, one in a finite region of the conformal boundary, see Fig. \bref{figure}. Then, the geodesic length \eqref{geodesic} is given by
\be
\label{4pt_length}
\lengthB(w,\bar w) = \rads \left(\log w + \log \bar w\right) -\rads\log\varepsilon\;,
\ee
and comparing with \eqref{4pt_fin} we find that in this case the holomorphic block/length relation is
\be
\label{bl4}
f(w|\alpha, \beta,\epsilon_2) \sim -\frac{\;\epsilon_2}{\rads}\; \lengthB(w)\;,
\ee
where $\sim$ means up to constant and divergent contributions. The same relation holds in the global coordinates, where the both sides are
vanishing, see the last relation in \eqref{4pt_fin}.

\section{Concluding remarks: more than three background operators}
\label{sec:conclusion}

By way of conclusion, let us briefly outline  a  possible generalization of  the previous results. The 3-point and 4-point functions with
respectively two and three  background insertions  belong to the general family of  $n$-point large-$c$ functions with $n-k$ background
operators. Let us denote them as H$^{n-k}$L$^{k}$ type functions, where a true parameter is the number of perturbative operators $k$ measuring
a deviation form the classical $n$-point block. The previously considered cases are, therefore,  type H$^2$L and $H^3L$ functions.

Let AdS$_3[n-k]$ be a three-dimensional space with \banados defined by the classical energy-momentum tensor $ T(z|{\bf z})$ with $n-k$
singular points. The boundary Schwarz mappings and the Poincare coordinates  are build using the solutions of the associated Fuchsian
equation,
\be
\label{Tz}
\left[\frac{d^2}{dz^2}  + T(z|{\bf z})\right]\psi(z) = 0\;,
\quad
\text{where}
\quad T(z|{\bf z}) = \sum_{i=k+1}^n \frac{\epsilon_i}{(z-z_i)^2} + \frac{c_i}{z-z_i}\;,
\ee
where ${\bf z} = (z_{k+1},...,z_{n})$ are locations of the background operators with classical dimensions $\epsilon_i$, the $c_i$ are
respective accessory parameters. The resulting space AdS$_3[n-k]$ will have $n-k$ conical defects parameterized by background conformal
dimensions as can be directly seen from the Schwarz map of the $(z,\bar z)$-plane to some curvilinear polygon with $n-k$ vertices on the
$(w,\bar w)$-plane.

Assuming that $\epsilon_j/\epsilon_i \ll1$ for $j=1,...,k$ and $i = k+1, ..., n$ we can use the heavy-light approximation and introduce type
$H^{n-k}L^k$ perturbative conformal blocks $f_{(k,n-k)}(w)$. The point is that when we calculate such perturbative blocks using the monodromy
method,  the energy-momentum tensor arising in the zeroth order is exactly \eqref{Tz}.

On the other hand, within the monodromy method  the zeroth-order Fuchsian solutions are the degenerate $(n-k+1)$-point  conformal blocks of
the background operators taken in the  large-$c$ limit, where the additional operator is the degenerate light $\Psi_{(1,2)}$, see Section
\bref{sec:zeroth}. Thus, the auxiliary equation \eqref{Tz} responsible for the Schwarz mappings is the BPZ equation describing the degenerate
background blocks $\psi(z)$.

It is tempting to conjecture that  type $H^{n-k}L^k$ conformal blocks are equal to the length of  dual Steiner trees  in AdS$_3[n-k]$,
\be
\label{bl_gen}
f_{(k,n-k)}(w|\epsilon) \sim -\frac{1}{\rads}\; \lengthG (w|\epsilon)\;,
\ee
where the right-hand side is the weighted  length of the dual tree, and $w$ are locations of perturbative operators in the Poincare
coordinates which can be found from the general formula \eqref{roberts}.

\vspace{5mm}
\noindent \textbf{Acknowledgements.} We are grateful to V. Belavin for useful exchanges.  The work was supported by the Russian Science
Foundation grant 18-72-10123.

\appendix

\section{The lower-point conformal blocks: various details}
\label{app:A}

Here, we discuss  3-point and 4-point conformal blocks in the context of the heavy-light expansion.  All block functions are normalized as
$\cF \sim z^{\gamma}$ at $z\to 0$, where $\gamma$ is the linear combination of conformal dimensions. The other normalization $\cF \sim 1$ is
also possible, in which case, e.g. the decomposition \eqref{bfc} would explicitly contain $z^\gamma$ factors.

\paragraph{Conformal maps.} Under the map $z \rightarrow w(z)$ the $n$-point correlation functions of primary operators  transform as (see
e.g. \cite{DiFrancesco:1997nk})
\be
\langle \cO_1(w_1) \cdots  \cO_n(w_n)   \rangle = \prod^{n}_{i=1}\left(\frac{dw}{dz}\right)^{-\Delta_{i}}_{w=w_{i}}\left(\frac{d\bar w}{d\bar
z}\right)^{-\bar\Delta_{i}}_{\bar w=\bar w_{i}} \langle \cO_1(z_1) \cdots \cO_n(z_n)   \rangle\;.
\ee
It follows that large-$c$ perturbative conformal block with $k$ perturbative  and $n-k$  background  operators transforms as
\be
\label{class_tr}
f(w_1(z), ... ,w_k(z)|\epsilon, \tilde\epsilon) = f(z_1,...,z_k|\epsilon, \tilde\epsilon) - \sum_{i=1}^k\epsilon_i \log \frac{dw_i(z)}{dz_i}
\;,
\ee
where $\epsilon, \tilde\epsilon$ are classical external and intermediate dimensions of the perturbative operators.

\paragraph{3-point conformal block.}  In this case we have
\be
\label{3}
\langle \cO_2(z,\bar z)\cO_2(1) \cO_3(\infty)\rangle  = (\infty)^{2\Delta_3}C_{123} \, \cF(z) \bar{\cF}(\bar{z})\;,
\ee
where the (holomorphic) block can be defined as
\be
\label{3ptbl}
\cF(z) = (1-z)^{-(\Delta_1+\Delta_2-\Delta_3)}\;.
\ee
The block function at arbitrary $z_2$ and $z_3$ can be similarly defined.  Modulo infinite prefactors the correlation function  \eqref{3} can
be exponentiated to yield  the  3-point {\it classical} conformal block
\be
\label{3ptcl}
f(z|\epsilon) = -(\epsilon_1+\epsilon_2-\epsilon_3) \log(1-z)\;,
\ee
where the classical dimensions are $\epsilon_i = 6\Delta_i/c$. The 3-point function is exact within the heavy-light expansion.

\paragraph{4-point conformal block.}
Here  we reproduce  the perturbative block \eqref{thfb} directly from the original Virasoro block in the large-$c$  and the heavy-light
approximations. The $s$-channel block function can be  expanded near $z=0$ as \cite{Belavin:1984vu}
\be
\label{se}
\cF(z|\Delta_i, \tdelta, c) = z^{\tdelta - \Delta_1 - \Delta_2} \sum_{N=0}^{\infty} F_N z^N \;,
\ee
where the expansion coefficients $F_N = F_N(\Delta_i, \tdelta, c)$ in the lowest orders are given by
\be
F_0 = 1 \;,
\qquad
F_1 = \frac{(\tdelta - \Delta_1 + \Delta_2)(\tdelta - \Delta_4 + \Delta_3)}{2 \tdelta }\;,
\ee
\be
\ba{c}
\dps F_2 = \frac{( \tdelta +\Delta_2 - \Delta_1) ( \tdelta + \Delta_2 - \Delta_1 + 1) ( \tdelta + \Delta_3 - \Delta_4) ( \tdelta + \Delta_3 -
\Delta_4 + 1)}{4  \tdelta (2  \tdelta + 1)} + \\
\\
\dps + 2\left(\frac{\Delta_1 + \Delta_2}{2} + \frac{3 (\Delta_1 - \Delta_2)^2}{2(1 + 2 \tdelta)} + \frac{( \tdelta -1) \tdelta}{2 (1 + 2
\tdelta)}\right) \left(c + \frac{2 \tdelta ( 8 \tdelta - 5)}{(1 + 2 \tdelta)}\right)^{-1}\times \\
\\
\dps \times \left(\frac{\Delta_3 + \Delta_4}{2} +  \frac{3 (\Delta_4 - \Delta_3)^2}{2(1 + 2 \tdelta)} + \frac{( \tdelta - 1) \tdelta}{2 (1 + 2
\tdelta)}\right)\;.
\ea
\ee

\noindent To calculate perturbative classical blocks, we do the following steps.\footnote{Perturbative conformal block in the plane CFT were
considered in \cite{Hijano:2015rla,Alkalaev:2015wia,Alkalaev:2015lca,Alkalaev:2015fbw,Belavin:2017atm}. The same logic also applies to
(super)torus blocks with a heavy channel  along the non-contractible cycle
\cite{Alkalaev:2016ptm,Alkalaev:2016fok,Alkalaev:2017bzx,Alkalaev:2018qaz}.}
\begin{itemize}

\item The large-$c$ regime, the classical dimensions:
\be
\epsilon_{i,p} = 6\Delta_{i,p}/c\;,
\qquad
i=1,2,3,4\;.
\ee

\item The classical block (see \eqref{class}):
\be
f(z|\epsilon_i, \epsilon_p)  \approx \frac{6}{c}\log \cF(z|\Delta_i, \tdelta, c) +\cO(1/c)\;.
\ee

\item The lightness parameter $\delta\ll1$: rescale a part of conformal dimensions as $\epsilon_{i,p} \rightarrow \delta \epsilon_{i,p}$ and
    expand $f= f(z|\epsilon_i, \epsilon_p)$ in powers of $\delta$,
\be
f =\frac{1}{\delta^{s}} f_{-s} +\frac{1}{\delta^{s-1}} f_{-s+1}+...+ \delta^0 f_0+ \delta f_{1} + \cO(\delta^2)\;,
\ee for some $s\in \mathbb{N}$. Since for general dimensions we have  the Laurent series, we require that all singular terms $f_{-n}$ must
be vanishing that imposes constraints on conformal dimensions.  (For $n$-point blocks with two background operators this is $\epsilon_3 =
\epsilon_4$, for the 4-point block with three background operators  this is $\epsilon_1 = \epsilon_p$). The term $f_0$ is the leading
(background) classical block. The perturbative block is defined to be the first correction $f_{1}$.

\end{itemize}

Having three heavy background operators we rescale $\epsilon_2 \to \delta \epsilon_2$ and find the perturbative  block
\be
\label{me}
f_1= - \epsilon_2 \log z + \epsilon_2\sum_{N=1}^{\infty}  {\text f}_N z^N \;,
\ee
where
\be
\label{sep}
{\text f}_1 = \frac{1}{2}\;,
\qquad
{\text f}_2 = \frac{3 + 8 \epsilon_1 + 16 \epsilon_4}{8(3 + 4 \epsilon_1)}\;.
\ee
Note that rescaling further  $\epsilon_1 \rightarrow \delta \epsilon_1, \epsilon_p \rightarrow \delta \epsilon_p $ and decomposing again in
powers of $\delta$, we get the coefficients of the perturbative HHHL block  with $\epsilon_1 = \epsilon_p$ \cite{Hijano:2015rla}.

On the other hand, the small-$z$ expansion (modulo logarithms) of the perturbative block \eqref{thfb} is given by the same \eqref{me}, where
the first coefficients are given by
\be
{\text f}_1 = \frac{1}{2}\;,
\qquad
{\text f}_2 = \frac{2 \beta^2 + 4 \alpha^2 - 9}{8(\beta^2 - 4)}\;.
\ee
Using the change \eqref{ab} we can see that these coefficients coincide with \eqref{sep}.

\providecommand{\href}[2]{#2}\begingroup\raggedright\endgroup

\end{document}